\PassOptionsToPackage{usenames, dvipsnames, table}{xcolor}
\documentclass[10pt,conference]{IEEEtran}

\usepackage[breaklinks,colorlinks, urlcolor=black, linkcolor=black, citecolor=black]{hyperref}
\usepackage{xcolor}
\usepackage{color}
\usepackage{amsmath,amsopn}
\usepackage[ruled,boxed,commentsnumbered, linesnumbered]{algorithm2e}
\usepackage{subfigure}
\usepackage{endnotes,microtype,xspace,graphicx,fancyvrb,multirow}
\usepackage{booktabs}
\usepackage{array,underscore,relsize}
\usepackage[T1]{fontenc}
\usepackage{fancyhdr,totpages}
\usepackage{enumitem}
\usepackage[labelfont=bf,compatibility=false,font=small, skip=5pt]{caption}
\usepackage{multicol}

\usepackage{fp}
\usepackage{siunitx}
\usepackage{authblk}

\usepackage{listings}
\definecolor{lightgray}{rgb}{0.92, 0.92, 0.92}

\usepackage{balance}

\usepackage{verbatim}

\usepackage{algorithm2e} 
\usepackage{algpseudocode}
\usepackage[most]{tcolorbox}
\usepackage{footmisc}

\usepackage[square,comma,numbers,sort&compress]{natbib}

\newcommand{\algcomment}[1]{\tcp*[h]{\textcolor{BrickRed}{\scriptsize{#1}}}}

\fvset{fontsize=\scriptsize,xleftmargin=8pt,numbers=left,numbersep=5pt}

\def\Snospace~{\S{}}

\newif\ifdraft\drafttrue
\newif\ifnotes\notestrue
\ifdraft\else\notesfalse\fi

\newcommand{\eg}{{\em e.g.}}
\newcommand{\etal}{{\em et al.}}
\newcommand{\etc}{{\em etc.}}
\newcommand{\ie}{{\em i.e.}}

\input{glyphtounicode}
\newcolumntype{R}[1]{>{\raggedleft\let\newline\\\arraybackslash\hspace{0pt}}p{#1}}

\newcommand{\squishlist}{
\begin{itemize}[noitemsep,nolistsep,leftmargin=15pt]
\setlength{\itemsep}{-0pt}
}
\newcommand{\squishend}{
\end{itemize}
}

\usepackage{tikz}

\usepackage{bbding}
\usepackage{pifont}
\newcommand{\XV}{\ding{52}\textsuperscript{\kern-0.4em\ding{56}}\xspace}

\usepackage{xstring}
\newcommand{\PP}[1]{
\vspace{2px}
\noindent{\bf \IfEndWith{#1}{.}{#1}{#1.}}
}

\newcommand{\boxbeg}{
\vspace{2px}
\noindent\begin{tabular}{|l|}\hline
\begin{minipage}{3.2in}
\vspace{2px}
\noindent
}

\newcommand{\boxend}{
\vspace{2px}
\end{minipage}\\ \hline
\end{tabular}
}

\newcommand{\mybox}[2]{%
    \begin{tcolorbox}[colback=#1!5!white, boxrule=0.25mm, sharp corners, colframe=black, top=0pt, bottom=0pt, left=0pt, right=0pt]
        #2 %
    \end{tcolorbox}
}

\begin{document}

\title{Explainer-guided Targeted Adversarial Attacks against Binary Code Similarity Detection Models}

\author[1$\dag$]{Mingjie Chen}
\affil[1]{Zhejiang University}

\author[2$\dag$]{Tiancheng Zhu}
\affil[2]{Huazhong University of Science and Technology}

\author[3, 4]{Mingxue Zhang}
\affil[3]{The State Key Laboratory of Blockchain and Data Security, Zhejiang University}
\affil[4]{Hangzhou High-Tech Zone (Binjiang) Institute of Blockchain and Data Security}

\author[5]{Yiling He}
\affil[5]{University College London}

\author[6]{Minghao Lin}
\affil[6]{University of Southern California}

\author[7]{Penghui Li}
\affil[7]{Columbia University}

\author[3]{Kui Ren}

\date{}

\maketitle

\begingroup
\renewcommand\thefootnote{$\dag$}
\footnotetext[1]{The first two authors contributed equally to this work. Tiancheng Zhu conducted the research during his internship at Zhejiang University.}
\endgroup

\begin{abstract}
Binary code similarity detection (BCSD) serves as a fundamental technique for various software engineering tasks, \eg, vulnerability detection and classification.
Attacks against such models have therefore drawn extensive attention, aiming at misleading the models to generate erroneous predictions.  
Prior works have explored various approaches to generating semantic-preserving variants, \ie, adversarial samples, to evaluate the robustness of the models against adversarial attacks.
However, they have mainly relied on heuristic criteria or iterative greedy algorithms to locate salient code influencing the model output, failing to operate on a solid theoretical basis.
Moreover, when processing programs with high complexities, such attacks tend to be time-consuming.

In this work, we propose a novel optimization for adversarial attacks against BCSD models. In particular, we aim to improve the attacks in a challenging scenario, where the attack goal is to limit the model predictions to a specific range, \ie, the \emph{targeted attacks}. %
Our attack leverages the superior capability of black-box, model-agnostic explainers in interpreting the model decision boundaries, thereby pinpointing the critical code snippet to apply semantic-preserving perturbations.
The evaluation results demonstrate that compared with the state-of-the-art attacks, the proposed attacks achieve higher attack success rate in almost all scenarios, while also improving the efficiency and transferability. %
Our real-world case studies on vulnerability detection and classification further demonstrate the security implications of our attacks, highlighting the urgent need to further enhance the robustness of existing BCSD models.
\end{abstract}

\IEEEpeerreviewmaketitle

\section{Introduction}
\label{s:intro}
Binary code similarity detection (BCSD) measures the semantic similarity between binary functions by computing their similarity score.
BCSD is a foundational method that has supported many software engineering and security tasks, including vulnerability detection, malware analysis, software plagiarism detection, and binary code search~\cite{shimmi2024vulsim, chen2023deuedroid, jiang2024binaryai, luo2017semantics}.
It is also used in reverse engineering, patch analysis, deobfuscation, and cross-architecture code matching.

However, the increasing reliance on the BCSD models also raises concerns about their robustness against diverse attacks. 
Adversarial attacks, where the input function samples are carefully perturbed to mislead the model predictions, have emerged as a significant threat to the reliability of many deep learning models. 
Prior works have primarily %
used brute-force methods or heuristic rules for selecting the perturbation locations. 
For instance, brute-force approaches might iteratively remove each instruction from a binary program and select those that change the similarity score most~\cite{capozzi2024adversarial, capozzi2024lack}.
While potentially thorough, these attacks are fundamentally inefficient for practical application. 
On the other hand,  heuristic methods, such as selecting important instructions based on the frequency of their basic block's appearance on all execution paths~\cite{jia2022funcfooler}, may offer some improved efficiency but often yield suboptimal results and lack precision, primarily due to their reliance on pre-determined rules and insufficient consideration of complex inter-instruction relationships.

These studies demonstrate good performance primarily on \emph{untargeted attacks}, where the objective is to reduce the similarity score between two semantically similar functions. 
In contrast, \emph{targeted adversarial attacks}—which aim to mislead the model into assigning a high similarity score between an adversarial sample and a specific, semantically unrelated target function—have achieved limited success.
For example, the success rate of the state-of-the-art~\cite{capozzi2024lack} drops from 97.04\% to 45.4\% when moving from untargeted to targeted attacks, as mentioned in its evaluation.
Targeted attacks represent more practical threat scenarios in BCSD because they align closely with real-world adversarial goals that exploit model behavior in an intentional manner.
For instance, they enable adversaries to cloak plagiarized code, camouflage malware to resemble benign software, or misattribute known vulnerabilities to unrelated functions. %
However, targeted adversarial attacks are inherently more challenging to execute. 
Unlike untargeted attacks that only need to push an input across any nearby decision boundary to cause a misclassification, targeted attacks require meticulous control to guide it to a pre-selected incorrect region.

The recent advancement of explanation techniques, aiming to quantify the importance of input features to the model predictions, opens up a new possibility for optimizing the adversarial attacks.
Such techniques, commonly referred to as \emph{explainers} are typically used to generate saliency maps that highlight which parts of the input most influence the model’s output.
Prior work has applied explainers in the context of backdoor attacks~\cite{severi2021explanation}, where models are maliciously trained on poisoned training data to behave incorrectly on inputs containing specific triggers. 
Nonetheless, it remains unclear how explanations can be used to guide the adversarial attacks without modifying the model or its training process.

In this work, we aim to design and implement an explanation-guided optimization for targeted adversarial attacks against BCSD models.
Specifically, by leveraging explainers to pinpoint salient instructions within a binary sample, we systematically identify the most vulnerable code for perturbations, thereby improving the effectiveness and efficiency in adversarial sample generation.
However, precisely explaining the BCSD model predictions and generating the adversarial samples accordingly, is not trivial.
First, the explainers only estimate the importance of input \textit{features}, whereas the adversarial samples need to be constructed by perturbing the \textit{instructions}.
We need to precisely map the features to the specific instructions, which is particularly difficult for BCSD models that extract features on the basic block level (\emph{C1}).
Second, explainers identify salient instructions by analyzing the relationship between a pair of input functions. 
However, in the context of a targeted adversarial attack, multiple target functions can be involved.
To maximize the optimization benefits from explanation while minimizing the computational overhead, it is essential to strategically select the function pairs for explanation (\emph{C2}).

To solve the above challenges, we designed customized explanation generation approaches for four representative BCSD models in different architectures. 
Specifically, to address \emph{C1}, we implemented a sequence-based and graph-based instruction mapping strategy, to calculate the instruction importance based on the weights of features in different granularity.
To address \emph{C2}, we designed an iterative greedy algorithm, by choosing the least similar target function to the adversarial sample as the next explanation target.
This allows us to iteratively refine the adversarial sample to better mislead the target models.

We have evaluated our attacks on the binary functions from 8 real-world projects under four different compilation settings.
The results demonstrate that our method achieves higher success rates than the state-of-the-art attacks, with very little additional knowledge about the target models. 
Compared with the iterative instruction selection approach, our explanation-guided strategy effectively speeds up the instruction selection by up to 12.71x.
The experiments on two representative real-world security tasks, vulnerability detection and classification, further prove the real-world implications of our attacks.

In summary, we make the following contributions.
\squishlist
\item \PP{Explanation-guided Optimization} 
To the best of our knowledge, we are the first to leverage explainers for guiding adversarial attacks against BCSD models.
\item \PP{New Explanation Strategies}
We developed new strategies that can better explain the decision boundaries for both sequence- and graph-based BCSD models by mapping the feature space to the input instruction space.

\item \PP{Extensive Evaluation} 
We demonstrated that our attacks exhibit high effectiveness against existing BCSD models while maintaining high efficiency. %
\item \PP{Real-world Security Implications}
We showed how our attacks could successfully mislead BCSD models in vulnerability detection and classification.
\squishend

\section{Background}
\label{s:background}

\subsection{Binary Code Similarity Detection}
\label{s:background-bcsd}
Binary code similarity detection techniques (BCSD) are designed to identify similarities between binary programs or functions, \eg, when they are compiled from the same source using different compilers or in different optimization levels.
They serve as the fundamental solution to many problems, \eg, malware variant identification, software component analysis, and plagiarism detection~\cite{han2013malware, wagener2008malware, jiang2024binaryai, luo2017semantics}.
As syntactic-based detection can be easily bypassed, \eg, using code obfuscation, extensive efforts have been invested to identify the semantic similarities in binary code.
BCSD models generally operates in two main paradigms: sequence-based and graph-based.
Sequence-based models represent binary code as linear sequence of instructions (\eg, assembly code sequences),
and leverage string matching, or machine learning models to quantify the similarities~\cite{ding2019asm2vec, massarelli2021function, pei2022learning}. 
Graph-based models, in contrast, model binary code as graphs, \eg, control flow graphs (CFGs), abstract syntax trees, or other intermediate representation graphs~\cite{xu2017neural, yang2023asteria, li2019graph}.
They then employ graph embedding techniques or graph neural networks to capture the structural and contextual characteristics.

\subsection{Adversarial Attacks against BCSD Models}
\label{s:background-adv}
Adversarial attacks, aiming at misleading models to generate incorrect predictions, have been a critical research direction. 
These attacks exploit vulnerable models by subtly altering the input in ways that are often imperceptible to humans but can lead to erroneous outputs. 
Many critical applications, such as facial recognition, have come under scrutiny.

In addition to the image processing models, BCSD models are also recognized as fruitful targets for adversarial attacks~\cite{capozzi2024lack, capozzi2024adversarial, song2022mab, jia2022funcfooler, pierazzi2020intriguing, kreuk2018adversarial, lucas2021malware, kolosnjaji2018adversarial}. %
The goal of the adversary is to perturb a query function $f_Q$ into a semantically equivalent form, tricking the models into generating imprecise decisions.
More specifically, in targeted adversarial attacks, an adversary aims to maximize the similarity between $f_Q$ and a set of target functions, while in untargeted attacks, the goal is to minimize the similarity between $f_Q$ and its variants compiled from the same function.

One critical step in generating the adversarial samples is to choose the instructions on which semantic preserving perturbations can be applied.
To this end, prior works either rely on an brute-force traversal algorithm, or design customized heuristic rules.
For example, Capozzi \etal~\cite{capozzi2024lack} selects the perturbation locations by traversing the entire function and calculating the changes in similarity scores after removing each instruction.
Code perturbations like node splits are iteratively applied at the position of instructions that maximize similarity score changes.
Funcfooler~\cite{jia2022funcfooler}, on the other hand, relies on the heuristic rule that basic blocks in all execution paths from the function entry to exit are critical for model prediction.
Although effective, the brute-force and heuristic approaches are either computationally expensive, or unable to infer the intricate instruction importance, limiting the attack performance.

\subsection{Model Explainers}
The intricate architectures and nonlinear interactions of deep learning models significantly obscure the rationale behind model outputs, rendering the model decisions inaccessible to human intuition.
To address the problem, %
numerous model explainability frameworks
have been proposed to enhance the transparency.
Depending on criteria such as the prerequisite knowledge and application scenarios, the explainers can be categorized along two dimensions: white-box versus black-box, and task-specific versus model-agnostic, respectively. %

White-box explainers trace how the input features propagate through the network and contribute to the final decisions, \eg, by analyzing the gradients or activation patterns.
In contrast, black-box explainers aim to reason about model output without knowing the internal architecture nor parameters, which can be more practical in many security-sensitive scenarios.
On the other hand, task-specific explainers exploit the unique structure of a particular task (\eg, image classification, text generation) to generate explanations. 
For instance, when explaining image classification decisions, the explainers leverage domain knowledge like spatial hierarchies in images to highlight how the input influences the model output. 
Model-agnostic explainers work independently of the model architecture, making them universally applicable across any models such as decision trees, neural networks, and beyond.

\section{Target Models and Explainers}
\label{s:targets}

\subsection{Target Models}
\label{s:design-target}
For representativeness and diversity, we select two state-of-the-art sequence-based 
and graph-based 
BCSD models as our attack targets, respectively, which are also targeted in prior studies~\cite{capozzi2024lack, capozzi2024adversarial, jia2022funcfooler}.
In the following, we present the detailed design and especially, the feature space, of our target models.
Note that we implement the attacks using \emph{model-agnostic }and \emph{black-box} explainers.
Therefore, our proposed attacks can be extended to target other BCSD models with manageable engineering efforts, which we discuss in \autoref{s:discussion}.

\subsubsection{Sequence-based Models}
We select jTrans~\cite{wang2022jtrans} and SAFE~\cite{massarelli2021function} as the sequence-based target models. Both models extract features (\ie, embeddings) on the basis of instruction sequences.

\textbf{jTrans} extracts features from assembly instructions. It creates embeddings for each token (\ie, opcodes and operands) %
in an instruction using a pre-trained BERT model. 
The similarities of binary functions can then be measured based on the cosine similarities between normalized embedding vectors.

\textbf{SAFE} %
firstly performs preprocessing on the linear sequence of assembly instructions.
It replaces the memory addresses and immediate values that exceed a threshold with placeholder tokens, \eg, \texttt{MEM} and \texttt{IMM}.
It then maps each normalized instruction to a vector representation (embedding) 
using a Self-Attentive Neural Network, which combines a bi-directional GRU RNN with an attention mechanism and a final fully connected layer. %
The similarity between two functions consists of the cosine distance between their corresponding embeddings.

\subsubsection{Graph-based Models}
We select Gemini~\cite{xu2017neural} and GMN~\cite{li2019graph} as the graph-based target models, which extract features from graph representations of binary functions.

\textbf{Gemini} generates an attributed CFG (ACFG) to encode the features of an input function. Specifically, the ACFG retains the control structure while encoding the features of each basic block into an eight-tuple representation. %
A Siamese network model is then trained for generating embedding vectors of the ACFGs. 
Similarities between two functions can be measured using the cosine similarity between the embedding vectors of the ACFGs. 
Some important features within a basic block include $n_{con}$, which counts constant-type instructions (\eg{,} \texttt{add ax, 1}), 
$n_{str}$, which counts string-type instructions (\eg{,} \texttt{mov ax, [string\_address]}), 
$n_{trans}$, which counts transfer-type instructions (\eg{,} \texttt{mov}), 
$n_{call}$, which counts call instructions (\eg{,} \texttt{call}), 
and $n_{ari}$, which counts arithmetic instructions (\eg{,} \texttt{add}).

\textbf{GMN}
is a versatile graph matching model, \ie,
it imposes no constraints on features or structures of the graphs, providing flexibility for choosing arbitrary feature extraction strategies.
By employing an attention mechanism, GMN facilitates information exchange between two graphs, enabling each graph to simultaneously capture intra-graph node adjacency information, and inter-graph matching information. This dual-level information integration enhances the accuracy of embedding vectors by more effectively encoding structural and semantic relationships both within and across the graphs.
Finally, similarities between binary functions can be measured based on the Euclidean and cosine distance between the corresponding embedding vectors. %

\subsection{Explainers}
\PP{LIME}
As sequence-based models extract features based on instruction sequences.
In this work, we utilize LIME~\cite{ribeiro2016should} as the explainer for comprehending the model predictions. 
Specifically, LIME generates a local feature dataset around the input features of a target model, and then fits a linear model on the local dataset. The weights of the linear model reflect the importance of input features. 
We made the design choice because LIME is among the most representative explainers and is able to efficiently measure the importance of sequential features.
The model-agnostic nature also ensures high transferability of our attacks (\autoref{s:eval-transfer}).
Nonetheless, our attacks can also be implemented using other explainers like SHAP~\cite{NIPS2017-7062}, which we discuss in \autoref{s:discussion}.

\PP{GNNExplainer}
To generate accurate explanations for graph-based BCSD models, we employed GNNExplainer~\cite{ying2019gnnexplainer}.
Compared with LIME, which mainly fits a simple linear model, GNNExplainer is able to capture the complex relationships among components in the graph representations.
Specifically, GNNExplainer 
takes the adjacency matrix and a feature matrix as input, and 
iteratively updates an edge mask $Mask^{edge}_{L}$ and a node feature mask $Mask^{feature}_{(N, M)}$ to infer the feature importance. 
Here, $L$ and $N$ represent the number of edges and nodes in the graph representation of the input (\eg, the CFG), respectively.
$M$ represents the number of features extracted from each node. 
The goal is to make the predictions on the current subgraph as similar as possible to the original predictions. 
Through iterative mask optimization, different subgraphs are constructed and evaluated, until an ``optimal'' subgraph is  obtained. %
In this work, we utilize the final node feature mask to pinpoint important features.
Similarly, the design principle can be implemented using other graph-related explainers.
\section{Design}
\label{s:design}

\begin{figure}[!t]
    \center{\includegraphics[width=1\columnwidth]{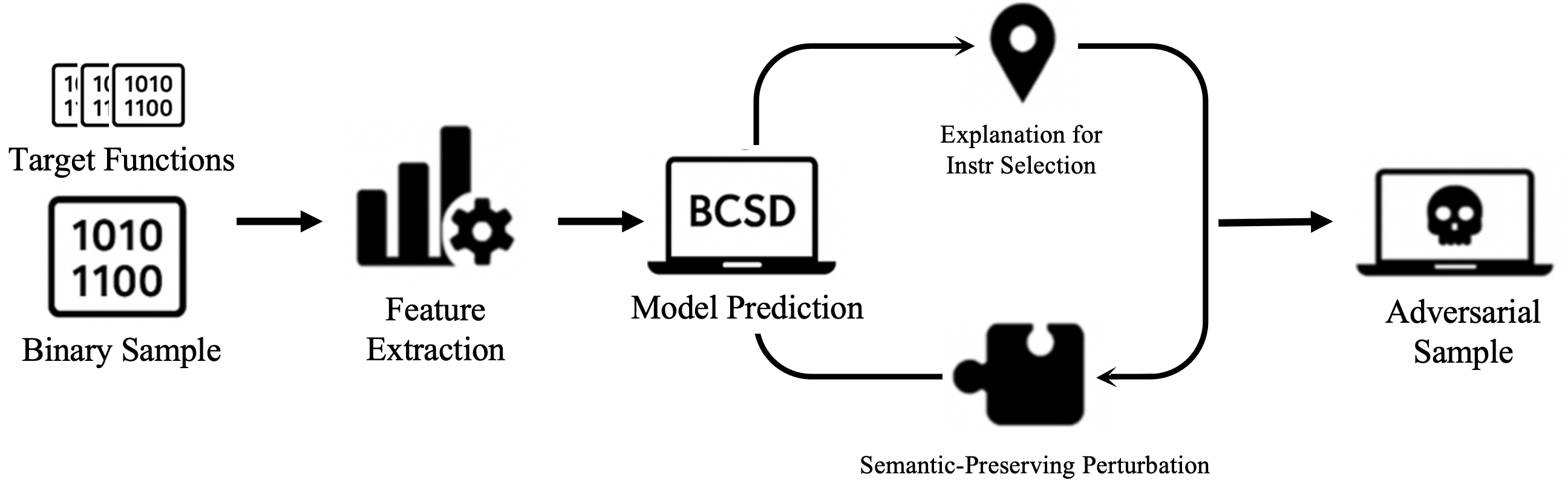}}
    \caption{The workflow of explainer-guided adversarial attacks.} %
    \label{fig:workflow}
\end{figure}

Inspired by the ability of explainers to pinpoint critical features that influence model predictions, 
in this work, we proposed to utilize explainers for guiding adversarial attacks against BCSD models.
This provides precise guidance for more efficient adversarial sample generation.
The workflow of our attacks is presented in \autoref{fig:workflow}. 
Given a binary sample and a list of target functions as input, the adversary firstly %
employs explainers to infer the importance of each feature.
She or he then correlates the important features with the corresponding instructions, applying semantic-preserving perturbations to generate an intermediate adversarial sample.
Based on the observed similarity changes, the adversary iteratively refines the adversarial samples by repeatedly performing explanations and perturbations until the termination conditions are satisfied.

In the following, we first describe our threat model in~\autoref{s:design-threat}. %
We then demonstrate how we generate explanations for different model predictions in \autoref{s:design-explain}, and how they guide the adversarial sample generation in \autoref{s:design-ae-generation}.
Finally, we provide the implementation details in \autoref{s:impl}.

\subsection{Threat Model}
\label{s:design-threat}
In this work, we aim to conduct targeted attacks against representative sequence-based and graph-based BCSD models.
We assume the adversary has limited internal knowledge about the target models, \ie, the extracted features and the categories of model architectures.
In the meanwhile, the adversary can query the target models to obtain arbitrary numbers of input-output pairs.
Note that such a threat model is practical and commonly adopted in various adversarial attack scenarios, \eg, attacks against image processing and similarity detection models~\cite{capozzi2024adversarial, biggio2018wild, lapid2023see}.

\subsection{Explanation Generation}
\label{s:design-explain}
\begin{figure*}[!t]
    \center{\includegraphics[width=\textwidth, trim=10pt 0 0 0, clip]{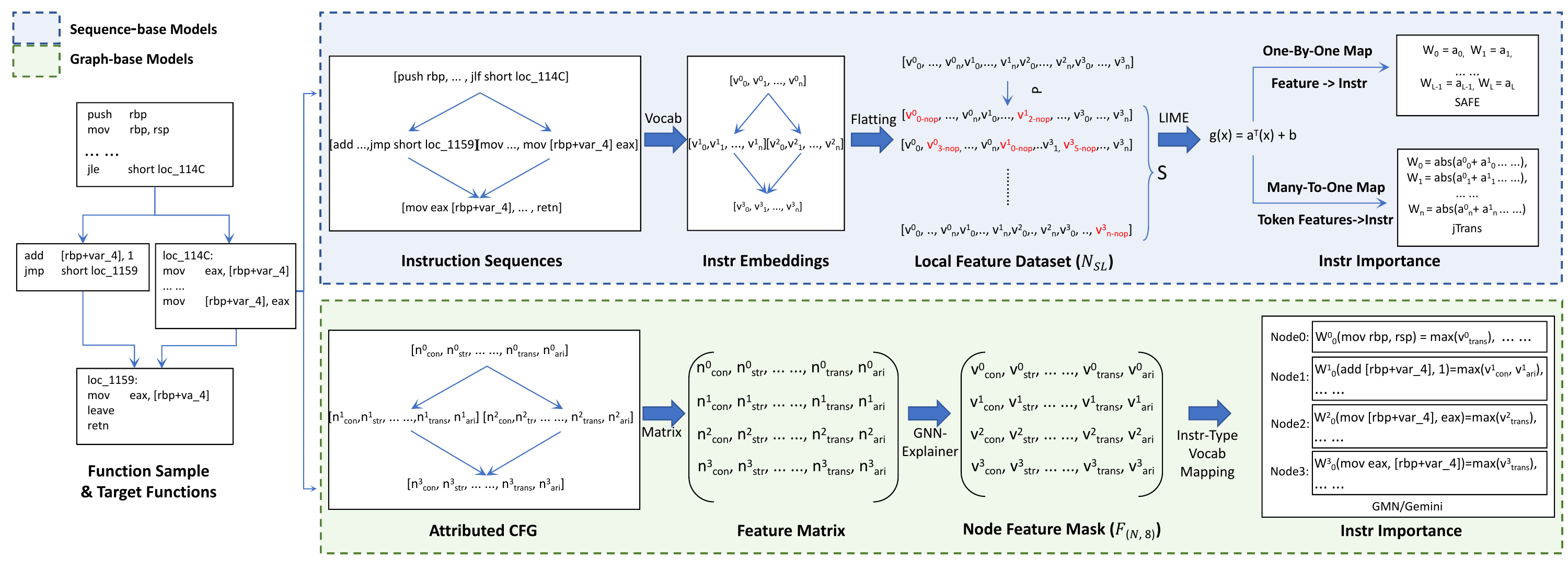}}
    \caption{Explanation generation for sequence-based and graph-based BCSD models.
    }
    \label{fig:LIME-JTRANS}
\end{figure*}

We now describe how we generate explanations for the predictions of target models,
and how we select the important instructions accordingly.
The detailed design is organized into two categories, as depicted in \autoref{fig:LIME-JTRANS}.

\subsubsection{Sequence-based Models}
\label{s:design-explain-sequence}

As described in \autoref{s:targets}, LIME approximates the decision boundary based on a local perturbed dataset.
Constructing the dataset, however, is not straightforward for our attacks. LIME by default perturbs the features of a large corpus of input, while retaining their original distribution.
When generating explanations for our attacks, the distribution of the original feature space is unknown.
Therefore, in this work, we modified LIME to generate a customized local feature dataset.
Suppose the input features are denoted as $N_L$, where $L$ %
represents the number of features extracted from a binary sample. 
We aim to generate a local feature dataset $N_{SL}$, which consists of $S$ perturbed variants of $N_L$.
Specifically, every $N_{SL}^i \in N_{SL}, i \in \{0, 1, ..., S-1\}$ is composed of $L$ perturbed features, each constructed by replacing $N_L^j, j \in \{0, 1, ..., L-1\}$ with the 
corresponding feature of \texttt{NOP} (no-operation instruction) with a probability $p$. %
We then feed $N_{SL}$ into LIME to approximate the decision boundary $f$ using a local linear model $g$, thereby determining the importance of each feature.
The optimization objective of LIME can be expressed as minimizing the following function:

\begin{equation}
\label{equ:opt-obj}
\sum_{k=0}^{S-1}\exp\left(-\frac{dis(\mathbf{N}_\mathrm{L},\mathbf{N}_\mathrm{SL}^{k})^2}{\sigma^2}\right)\left(f(\mathbf{N}_\mathrm{SL}^{k})-g(\mathbf{N}_\mathrm{SL}^{k})\right)^2 \,.
\end{equation}
\vspace{1mm}

Here, \( \sigma \) is a hyperparameter that controls the rate of weight decay.
The $dis$ function is used to calculate the distance between the perturbed features $N_{SL}^i$
and the initial features $N_L$.
Ultimately, we obtain a fitted linear function  \( g(x)=a^Tx +b \), where \( a \) represents the vector of corresponding weights.

Note that sequence-based models may derive multiple features out of each instruction.
In order to generate an adversarial sample, we need to locate the salient \textit{instructions} based on the explanation results, \ie, the importance of different \textit{features}.
In the following, we describe how the salient instructions are selected.

\PP{Explaining jTrans.} 
As stated above, we use LIME to infer the importance (weights) of each feature in the binary sample.
Since jTrans extracts features for each opcode and operand, we can calculate the importance of each instruction based on the explanations. 
Specifically, we aggregate the weights of tokens that correspond to each instruction by calculating the sum of the absolute values of such weights.
We then select the most critical instructions based on the aggregated weights. %

\PP{Explaining SAFE.} 
Different from jTrans, SAFE extracts one feature vector for each preprocessed instruction, as described in \autoref{s:design-target}.
Therefore, we can feasibly select the important instructions based on the weights derived from LIME.

\subsubsection{Graph-based Models}
\label{s:design-explain-graph}
Different from LIME, GNNExplainer does not require a perturbed dataset for learning the prediction boundaries.
However, as the graph-based models extract features on the basic block level, 
there obviously exists a gap between the explanation results and the importance of instructions.
To solve the problem, we designed an \textit{instruction-type vocabulary mapping} mechanism to calculate the importance of instructions, as illustrated below.

\PP{Explaining Gemini and GMN.} 
For an input function, Gemini first converts it into an ACFG, where each basic block is represented as an eight-tuple.
The ACFG can thus be encoded as a feature matrix $A_{(N,8)}$,
where $N$ represents the total number of basic blocks in the ACFG. 
Next, \( A_{(N,8)} \) is fed into GNNExplainer, which iteratively updates the mask matrices to obtain the final node feature mask \( F_{(N,8)} \) as the explanation. Here,  $F_{(N,8)}$ indicates the importance of each item in the feature tuples, \eg, $n_{cons}$ and $n_{ari}$.
In the instruction-type vocabulary mapping, we record the ``type'' for each instruction, \eg, constant-type, string-type, transfer-type, \etc, in a hash table.
The importance of instructions can thus be calculated as the weights of the features in the corresponding type.
For instance, we use the weight of $n_{ari}$ in basic block $B_i$ as the importance score of all arithmetic instructions in $B_i$.
Note that an instruction can be associated with multiple types and thus multiple feature weights, \eg, \texttt{add ax, 1} is a constant-type and arithmetic instruction.
In such cases, we use the maximum feature weight as the importance score of the instruction. 
Our evaluation results proved that such a design is simple yet effective (\autoref{s:eval}).
Finally, we can select the instructions with the highest importance scores as the candidates for applying perturbations.

As GMN is compatible with arbitrary features, in this work, we use the same features as in Gemini. Therefore, the salient instructions can be selected using the same way as stated above.
Nonetheless, GNNExplainer can also be used to guide the adversarial attacks against graph-based models using other features, which we discuss in \autoref{s:discussion}.

\subsection{Adversarial Sample Generation}
\label{s:design-ae-generation}
Our next goal is to generate an adversarial sample by applying semantic-preserving perturbations to the important instructions selected above.
The detailed algorithm is presented in \autoref{alg:adv-generation}, which
iteratively refines an adversarial sample, until the similarity between the input and target functions exceeds a predefined threshold, or the maximum number of iterations has been encountered (Line 3-17).
The goal is to maximize the minimum similarity between the adversarial sample with the target functions, which can be expressed as:
\vspace{1mm}
\begin{equation}
\label{equ:opt-obj}
\max_{F_s^k}\left(\min_{i \in \{0, 1, ..., m\}}\left(sim({F_s}^{k}, {F_t}^{i})\right)\right) \,.
\end{equation}
\vspace{1mm}

Here, $F_s^k$ denotes the intermediary adversarial sample generated in the $k$-th iteration, and $F_t^i$ represents one of the $m$ target functions.

Specifically, we first use explainers to calculate the weights of features, and select the important instructions accordingly (Line 4-5), as described in \autoref{s:design-explain}.
For each important instruction, we randomly select a list of candidate perturbations that can be applied  (Line 7).
Prior works have extensively studied the effect of various semantic-preserving instruction perturbations.
In this work, we utilize the existing perturbations (\ie, dead branch insertion, basic block split, instruction re-ordering, and equivalent instruction replacement), as it is not our main focus to design novel code perturbation methods.
However, the proposed attacks may indeed be optimized by integrating advanced program obfuscation techniques, which will be discussed in \autoref{s:discussion}.
Applying the selected perturbations, we then obtain a list of intermediary adversarial samples (Line 9).
With a probability of $p_u$, the most similar sample to the input function is updated as the final adversarial sample, helping to escape local optima (Line 10-12).

In particular, we adopt a greedy strategy for selecting the explanation targets, \ie, by choosing the 
least similar target function to our adversarial sample for explanation generation (Line 15).
Such a design helps to maximize the potential optimization from explanations, while also avoiding the computational cost to explain all combinations of target functions and the adversarial samples.

\SetAlCapNameFnt{\small}
\SetAlCapFnt{\small}
\begin{algorithm}[t]
    \caption{Adversarial binary sample generation with explanation guidance. 
    }
    \small
    \label{alg:adv-generation}
    \SetKwInOut{Input}{Input}
    \SetKwInOut{Output}{Output}
    \SetKwFunction{Explain}{Explain}
    \SetKwFunction{MapToInstr}{MapToInstr}
    \SetKwFunction{ApplyPerturbation}{ApplyPerturbation}
    \SetKwFunction{GetMinSimilarity}{GetMinSimilarity}
    \SetKwFunction{UpdateAdv}{UpdateAdv}
    \SetKwFunction{sim}{sim}

    \Input{Input function $F_s$, target functions $F_t = [F^0_t, F^1_t, \dots, F^m_t]$, perturbations $P$}
    \Output{An adversarial function $F_s^{\prime}$}

    $iter \gets 0$, \quad $F_t^c \gets \texttt{random}(F_t)$, \quad $maxSim \gets -1$ \\
    $F_s^{\prime} \gets F_s$ \algcomment{Initialize adversarial candidate}\\

    \While{$\sim(F_s^{\prime}, F_t^c) < thres$ \textbf{and} $iter < maxIter$}
    {
        $featureWeights \gets$ \Explain($F_s$, $F_t^c$) \algcomment{Compute feature importance}\\
        $candidateInstr \gets$ \MapToInstr($featureWeights$) \algcomment{Select candidate instructions}\\

        \ForEach{$instr \in candidateInstr$}
        {
            $instrP \gets \texttt{random}(P)$ \algcomment{Generate perturbation set}\\
            \ForEach{$P_c \in instrP$}
            {
                $F_s^P \gets$ \ApplyPerturbation($F_s$, $P_c$) \\
                \If{\sim($F_s^P$, $F_t^c$) $> maxSim$}
                {
                    $F_s^{\prime} \gets$ \UpdateAdv($F_s^P$, $p_u$) \algcomment{Update AS}
                }
            }
        }

        $F_t^c \gets$ \GetMinSimilarity($F_s^{\prime}$, $F_t$) \algcomment{Update target}\\
        $iter \gets iter + 1$
    }
    \Return $F_s^{\prime}$
\end{algorithm}

\subsection{Implementation}
\label{s:impl}
We implement our attacks based on PyTorch 1.6.0 with CUDA 10.2 and CUDNN 7.6.5. 
The control flow graphs of input functions are extracted using Radare2~\cite{radare2} and angr~\cite{angr}.
We run all experiments on a Linux server running Debian 12.2.0, with an Intel Xeon 6230 at 2.10GHz
with 80 virtual cores including hyperthreading, 503 GB RAM, and two Nvidia RTX 4090 GPU.
\section{Evaluation}
\label{s:eval}
In this section, we present a comprehensive evaluation of our explainer-guided adversarial attacks. 
In particular, we aim to answer the following research questions.

\squishlist
\item \textbf{Effectiveness (RQ1)}: Can the proposed attacks effectively mislead state-of-the-art BCSD models? 
\item \textbf{Efficiency (RQ2)}: To what extent can explainers improve the efficiency of targeted and untargeted adversarial attacks against BCSD models?
\item \textbf{Transferability (RQ3)}: Can the adversarial samples generalize to attack other BCSD models?
\item \textbf{Real-world Implications (RQ4)}: How effective are our attacks in real-world application scenarios, \eg, vulnerability detection and classification?
\squishend

\subsection{Experimental Setup}
\label{s:eval-setup}
\subsubsection{Baseline}
We carefully studied the literature for the baseline.
Among the three state-of-the-art adversarial attacks against BCSD models-$A_1$~\cite{capozzi2024adversarial}, $A_2$~\cite{capozzi2024lack} and $A_3$~\cite{jia2022funcfooler},
$A_2$ essentially extends $A_1$ by integrating more perturbation approaches on a wider range of BCSD models.
In the meanwhile, the implementation of $A_3$ is not publicly available, preventing a fair and reproducible comparison. 
Therefore, $A_2$ is the only available tool,\footnote{Though not publicly available, we obtained the source code from the authors for our comparison.}
and we selected it as the primary baseline for our comparative evaluation.

\subsubsection{Dataset}
\label{subsub:Dataset}
To conduct a fair comparison, we used the same dataset as in the baseline attack.
More specifically, the dataset consists of eight open-source projects written in C: binutils~\cite{binutils}, curl~\cite{curl}, openssl~\cite{openssl}, sqlite~\cite{sqlite}, gsl~\cite{gsl}, libconfig~\cite{libconfig}, ffmpeg~\cite{ffmpeg}, and postgresql~\cite{postgresql}.
The projects were compiled on Ubuntu 20.04, using two compilers (\ie, gcc-9.4.0 and clang-12) and two different optimization levels (\ie, O0 and O3).
Therefore, each program is generated under four compilation configurations.

\subsubsection{Assessment Criteria}
We now define the assessment criteria for measuring the performance of our proposed attacks.

\PP{Attack Success Rate (ASR)}
We calculate the attack success rate to measure the effectiveness of our attacks.
As described in \autoref{s:design-ae-generation}, in a targeted adversarial attack, an adversary aims to maximize the similarity between an input function and a set of target functions.
Therefore, we define the attack success rate (ASR@i) as the percentage of attacks that successfully bring \textit{i} target functions to the top-K most similar functions in a function pool.
Similar to the baseline approach, we define an aggregated success rate \textbf{wASR} as 0.25*ASR@1 + 0.5*ASR@2 + 0.75*ASR@3 + ASR@4.

\PP{Required Modifications (M-Instrs and M-Nodes)}
We additionally measured the number of instructions and basic blocks injected in the adversarial samples, after the iterative refinement process.

\PP{Overhead}
To quantitatively measure the efficiency, especially the speedup achieved by using the explainers,
we record the execution time required for selecting important instructions and generating the adversarial samples.

\subsubsection{Parameter Selection}
We adopted the default recommended configuration of all target models in the experiments.
Similar to the baseline attack, we set the similarity threshold for early termination of the iterative refinement as 1.0, and
the maximum number of iterations as 30.
For each candidate instruction, we randomly sampled 2,100 candidate perturbations and selected the sub-optimal adversarial sample with a probability of 0.1.

\begin{table*}[t]
\label{tab:result-untargeted}
    \caption{Evaluation results of our attacks. M+ and M represent the explainer-guided and baseline attacks against model M, respectively. ``INIT'' refers to the attack success rates using the original input sample. The results are based on the top-5 functions returned by each model.}
\resizebox{\textwidth}{!}{
\begin{tabular}{c|c|llllllll|llllllll}
\toprule
\multicolumn{1}{l|}{}          & \multicolumn{1}{l|}{\multirow{2}{*}{}} & \multicolumn{8}{c|}{\textbf{Sequence-based Models}}                                                                                                                                                                                                                              & \multicolumn{8}{c}{\textbf{Graph-based Models}}                                                                                                                                                                                                                                  \\ \cmidrule{3-18} 
\multicolumn{1}{l|}{\textbf{}} & \multicolumn{1}{l|}{}                  & \multicolumn{2}{c|}{\textbf{Safe+}}                                     & \multicolumn{2}{c|}{\textbf{Safe}}                                      & \multicolumn{2}{c|}{\textbf{jTrans+}}                                     & \multicolumn{2}{c|}{\textbf{jTrans}}                 & \multicolumn{2}{c|}{\textbf{Gemini+}}                                   & \multicolumn{2}{c|}{\textbf{Gemini}}                                    & \multicolumn{2}{c|}{\textbf{GMN+}}                                      & \multicolumn{2}{c}{\textbf{GMN}}                   \\ 
\cmidrule{2-18} 
\textbf{}                      & \textbf{|P|}                           & \multicolumn{1}{c}{\textbf{128}} & \multicolumn{1}{c|}{\textbf{512}} & \multicolumn{1}{c}{\textbf{128}} & \multicolumn{1}{c|}{\textbf{512}} & \multicolumn{1}{c}{\textbf{128}} & \multicolumn{1}{c|}{\textbf{512}} & \multicolumn{1}{c}{\textbf{128}} & \multicolumn{1}{c|}{\textbf{512}} & \multicolumn{1}{c}{\textbf{128}} & \multicolumn{1}{c|}{\textbf{512}} & \multicolumn{1}{c}{\textbf{128}} & \multicolumn{1}{c|}{\textbf{512}} & \multicolumn{1}{c}{\textbf{128}} & \multicolumn{1}{c|}{\textbf{512}} & \multicolumn{1}{c}{\textbf{128}} & \multicolumn{1}{c}{\textbf{512}} \\ 
\midrule
\multirow{4}{*}{\textbf{@1}}   &  \multicolumn{1}{l|}{\textbf{INIT}}                & \multicolumn{1}{c|}{0.41}           & \multicolumn{1}{c|}{0.08}                & \multicolumn{1}{c|}{0.42}            & \multicolumn{1}{c|}{0.08}                & \multicolumn{1}{c|}{0.49}            & \multicolumn{1}{c|}{0.12}                & \multicolumn{1}{c|}{0.49}            & \multicolumn{1}{c|}{0.12}               & \multicolumn{1}{c|}{0.53}            & \multicolumn{1}{c|}{0.23}                & \multicolumn{1}{c|}{0.53}            & \multicolumn{1}{c|}{0.23}                & \multicolumn{1}{c|}{0.65}            & \multicolumn{1}{c|}{0.16}                & \multicolumn{1}{c|}{0.67}            & \multicolumn{1}{c}{0.17}                \\
                               & \multicolumn{1}{l|}{\textbf{ASR}}                   & \multicolumn{1}{c|}{\textbf{0.81}}            & \multicolumn{1}{c|}{\textbf{0.47}}                & \multicolumn{1}{c|}{0.78}            & \multicolumn{1}{c|}{0.42}                & \multicolumn{1}{c|}{\textbf{0.91}}            & \multicolumn{1}{c|}{\textbf{0.54}}                & \multicolumn{1}{c|}{0.82}            &  
                               \multicolumn{1}{c|}{0.44}              &
                               \multicolumn{1}{c|}{\textbf{0.67}}            & \multicolumn{1}{c|}{\textbf{0.28}}                & \multicolumn{1}{c|}{0.63}            & \multicolumn{1}{c|}{0.24}                & \multicolumn{1}{c|}{\textbf{0.79}}            & \multicolumn{1}{c|}{0.24}                & \multicolumn{1}{c|}{0.70}            &           \multicolumn{1}{c}{\textbf{0.25}} \\
                               & \multicolumn{1}{l|}{\textbf{M-Instrs}}    & \multicolumn{1}{c|}{209.09}            & \multicolumn{1}{c|}{201.43}                & \multicolumn{1}{c|}{206.92}            & \multicolumn{1}{c|}{206.10}                & \multicolumn{1}{c|}{103.67}            & \multicolumn{1}{c|}{112.54}                & \multicolumn{1}{c|}{83.44}            &            \multicolumn{1}{c|}{82.32}               & 
                               \multicolumn{1}{c|}{314.79}            & \multicolumn{1}{c|}{301.64}                & \multicolumn{1}{c|}{336.46}            & 
                               \multicolumn{1}{c|}{304.58}                & \multicolumn{1}{c|}{140.57}            & 
                               \multicolumn{1}{c|}{146.67}                & \multicolumn{1}{c|}{109.66}            &           \multicolumn{1}{c}{94.56}      \\
                               & \multicolumn{1}{l|}{\textbf{M-Nodes}}    & \multicolumn{1}{c|}{24.30}            & \multicolumn{1}{c|}{25.11}                & \multicolumn{1}{c|}{25.33}            & \multicolumn{1}{c|}{25.95}                & \multicolumn{1}{c|}{4.11}            & \multicolumn{1}{c|}{4.96}                & \multicolumn{1}{c|}{3.61}            &             \multicolumn{1}{c|}{3.77}                  & 
                               \multicolumn{1}{c|}{21.42}            & \multicolumn{1}{c|}{22.68}                & \multicolumn{1}{c|}{19.37}            & 
                               \multicolumn{1}{c|}{19.08}                & \multicolumn{1}{c|}{7.01}            & 
                               \multicolumn{1}{c|}{3.08}                & \multicolumn{1}{c|}{4.27}            &             \multicolumn{1}{c}{3.68}    \\ 
\midrule
\multirow{4}{*}{\textbf{@2}}   & \multicolumn{1}{l|}{\textbf{INIT}}                  & \multicolumn{1}{c|}{0.20}            & \multicolumn{1}{c|}{0}                & \multicolumn{1}{c|}{0.18}            & \multicolumn{1}{c|}{0}                & \multicolumn{1}{c|}{0.08}            & \multicolumn{1}{c|}{0.01}                & \multicolumn{1}{c|}{0.08}            & \multicolumn{1}{c|}{0.01}               & \multicolumn{1}{c|}{0.23}            & \multicolumn{1}{c|}{0.04}                & \multicolumn{1}{c|}{0.23}            & \multicolumn{1}{c|}{0.04}                & \multicolumn{1}{c|}{0.24}            & \multicolumn{1}{c|}{0.02}                & \multicolumn{1}{c|}{0.24}            & \multicolumn{1}{c}{0.02}                  \\
                               & \multicolumn{1}{l|}{\textbf{ASR}}            & \multicolumn{1}{c|}{\textbf{0.69}}            & \multicolumn{1}{c|}{\textbf{0.15}}                & \multicolumn{1}{c|}{0.67}            & \multicolumn{1}{c|}{0.13}                & \multicolumn{1}{c|}{\textbf{0.24}}            & \multicolumn{1}{c|}{\textbf{0.03}}                & \multicolumn{1}{c|}{0.15}            & 
                               \multicolumn{1}{c|}{0}               &
                               \multicolumn{1}{c|}{\textbf{0.27}}            & \multicolumn{1}{c|}{\textbf{0.10}}                & \multicolumn{1}{c|}{0.26}            & \multicolumn{1}{c|}{0.06}                & \multicolumn{1}{c|}{0.28}            & \multicolumn{1}{c|}{0.03}                & \multicolumn{1}{c|}{\textbf{0.31}}            &    
                               \multicolumn{1}{c}{\textbf{0.08}} \\
                               & \multicolumn{1}{l|}{\textbf{M-Instrs}}     & \multicolumn{1}{c|}{204.94}            & \multicolumn{1}{c|}{188.73}                & \multicolumn{1}{c|}{201.87}            & \multicolumn{1}{c|}{200.08}                & \multicolumn{1}{c|}{83.5}            & \multicolumn{1}{c|}{152.0}                & \multicolumn{1}{c|}{87.0}            &             \multicolumn{1}{c|}{-}          & 
                               \multicolumn{1}{c|}{291.04}            & \multicolumn{1}{c|}{262.80}                & \multicolumn{1}{c|}{289.81}            & 
                               \multicolumn{1}{c|}{248.50}                & \multicolumn{1}{c|}{144.61}            & 
                               \multicolumn{1}{c|}{189.67}                & \multicolumn{1}{c|}{118.65}            &           \multicolumn{1}{c}{178.38}      \\
                               & \multicolumn{1}{l|}{\textbf{M-Nodes}}   & \multicolumn{1}{c|}{24.96}            & \multicolumn{1}{c|}{28.27}                & \multicolumn{1}{c|}{25.97}            & \multicolumn{1}{c|}{27.08}                & \multicolumn{1}{c|}{4.92}            & \multicolumn{1}{c|}{6.0}                & \multicolumn{1}{c|}{4.0}            &              \multicolumn{1}{c|}{-}              & 
                               \multicolumn{1}{c|}{23.37}            & 
                               \multicolumn{1}{c|}{19.10}                & \multicolumn{1}{c|}{21.85}            & 
                               \multicolumn{1}{c|}{26.0}                & \multicolumn{1}{c|}{4.29}            & 
                               \multicolumn{1}{c|}{5.33}                & \multicolumn{1}{c|}{5.03}            &             \multicolumn{1}{c}{7.5}    \\ 
\midrule
\multirow{4}{*}{\textbf{@3}}   & \multicolumn{1}{l|}{\textbf{INIT}}                         & \multicolumn{1}{c|}{0.05}            & \multicolumn{1}{c|}{0}                & \multicolumn{1}{c|}{0.07}            & \multicolumn{1}{c|}{0}                & \multicolumn{1}{c|}{0}               & \multicolumn{1}{c|}{0}                & \multicolumn{1}{c|}{0}               & \multicolumn{1}{c|}{0}               & \multicolumn{1}{c|}{0.01}            & \multicolumn{1}{c|}{0.01}                & \multicolumn{1}{c|}{0.01}            & \multicolumn{1}{c|}{0.01}                & \multicolumn{1}{c|}{0.05}            & \multicolumn{1}{c|}{0}                & \multicolumn{1}{c|}{0.05}            & \multicolumn{1}{c}{0}                \\
                               & \multicolumn{1}{l|}{\textbf{ASR}}              & \multicolumn{1}{c|}{\textbf{0.43}}            & \multicolumn{1}{c|}{\textbf{0.04}}                & \multicolumn{1}{c|}{0.42}            & \multicolumn{1}{c|}{0.02}                & \multicolumn{1}{c|}{\textbf{0.02}}            & \multicolumn{1}{c|}{0}                & \multicolumn{1}{c|}{0.01}            & \multicolumn{1}{c|}{0}                  & \multicolumn{1}{c|}{\textbf{0.11}}            & \multicolumn{1}{c|}{\textbf{0.03}}                & \multicolumn{1}{c|}{0.05}            & \multicolumn{1}{c|}{0.02}                & \multicolumn{1}{c|}{0.05}            & \multicolumn{1}{c|}{0}                & \multicolumn{1}{c|}{\textbf{0.06}}            &             \multicolumn{1}{c}{\textbf{0.01}} \\
                               & \multicolumn{1}{l|}{\textbf{M-Instrs}}             & \multicolumn{1}{c|}{183.07}            & \multicolumn{1}{c|}{160.25}                & \multicolumn{1}{c|}{190.64}            & \multicolumn{1}{c|}{218.50}                & \multicolumn{1}{c|}{59.0}            & \multicolumn{1}{c|}{-}                & \multicolumn{1}{c|}{60.0}            &  
                               \multicolumn{1}{c|}{-}                & \multicolumn{1}{c|}{244.07}                 &    \multicolumn{1}{c|}{132.0}                & \multicolumn{1}{c|}{201.60}            & 
                               \multicolumn{1}{c|}{144.0}                & \multicolumn{1}{c|}{156.0}            & 
                               \multicolumn{1}{c|}{-}                & \multicolumn{1}{c|}{170.67}            &           \multicolumn{1}{c}{320.0}       \\
                               & \multicolumn{1}{l|}{\textbf{M-Nodes}}               & \multicolumn{1}{c|}{26.60}            & \multicolumn{1}{c|}{27.0}                & \multicolumn{1}{c|}{27.67}            & \multicolumn{1}{c|}{43.0}                & \multicolumn{1}{c|}{1.0}            & 
                               \multicolumn{1}{c|}{-}                & \multicolumn{1}{c|}{0}            &                 \multicolumn{1}{c|}{-}             & 
                               \multicolumn{1}{c|}{24.73}            &    \multicolumn{1}{c|}{25.33}                & \multicolumn{1}{c|}{29.60}            & 
                               \multicolumn{1}{c|}{30.0}                & \multicolumn{1}{c|}{3.2}            & 
                               \multicolumn{1}{c|}{-}                & \multicolumn{1}{c|}{4.67}            &             \multicolumn{1}{c}{4.0}     \\ 
\midrule
\multirow{4}{*}{\textbf{@4}}   & \multicolumn{1}{l|}{\textbf{INIT}}                          & \multicolumn{1}{c|}{0}            & \multicolumn{1}{c|}{0}                & \multicolumn{1}{c|}{0}            & \multicolumn{1}{c|}{0}                & \multicolumn{1}{c|}{0}            & \multicolumn{1}{c|}{0}                & \multicolumn{1}{c|}{0}            & \multicolumn{1}{c|}{0}               & \multicolumn{1}{c|}{0.01}         & \multicolumn{1}{c|}{0.01}                & \multicolumn{1}{c|}{0.01}         & \multicolumn{1}{c|}{0.01}                & \multicolumn{1}{c|}{0}            & \multicolumn{1}{c|}{0}                & \multicolumn{1}{c|}{0}            & \multicolumn{1}{c}{0}                 \\
                               & \multicolumn{1}{l|}{\textbf{ASR}}                 & \multicolumn{1}{c|}{\textbf{0.15}}            & \multicolumn{1}{c|}{0}                & \multicolumn{1}{c|}{0.13}            & \multicolumn{1}{c|}{0}                & \multicolumn{1}{c|}{0}            & 
                               \multicolumn{1}{c|}{0}                & \multicolumn{1}{c|}{0}            &   
                               \multicolumn{1}{c|}{0}                & \multicolumn{1}{c|}{\textbf{0.04}}            & \multicolumn{1}{c|}{\textbf{0.02}}                & \multicolumn{1}{c|}{0.02}            & \multicolumn{1}{c|}{0.01}                & \multicolumn{1}{c|}{\textbf{0.01}}            & \multicolumn{1}{c|}{0}                & \multicolumn{1}{c|}{0}            &   
                               \multicolumn{1}{c}{0}              \\
                               & \multicolumn{1}{l|}{\textbf{M-Instrs}}            & \multicolumn{1}{c|}{190.07}            & \multicolumn{1}{c|}{-}                & \multicolumn{1}{c|}{194.08}            & \multicolumn{1}{c|}{-}                & \multicolumn{1}{c|}{-}            &   
                               \multicolumn{1}{c|}{-}                & \multicolumn{1}{c|}{-}            &  
                               \multicolumn{1}{c|}{-}            &   
                               \multicolumn{1}{c|}{209.25}            &      \multicolumn{1}{c|}{133.0}                & \multicolumn{1}{c|}{144.0}            &     \multicolumn{1}{c|}{125.0}                & \multicolumn{1}{c|}{280.0}            &   \multicolumn{1}{c|}{-}                & \multicolumn{1}{c|}{-}            &                \multicolumn{1}{c}{-} \\
                               & \multicolumn{1}{l|}{\textbf{M-Nodes}}             & \multicolumn{1}{c|}{34.67}            & \multicolumn{1}{c|}{-}                & \multicolumn{1}{c|}{35.38}            & \multicolumn{1}{c|}{-}                & \multicolumn{1}{c|}{-}            &     \multicolumn{1}{c|}{-}                & \multicolumn{1}{c|}{-}            &                \multicolumn{1}{c|}{-}             &           \multicolumn{1}{c|}{33.0}            &      \multicolumn{1}{c|}{25.0}                & \multicolumn{1}
                               {c|}{30.0}            &   \multicolumn{1}{c|}{12.0}                & \multicolumn{1}{c|}{12.0}            &   \multicolumn{1}{c|}{-}                & \multicolumn{1}{c|}{-}            &                \multicolumn{1}{c}{-}  \\ 
 \bottomrule
\end{tabular}
}
\label{tab:asr-top5}
\end{table*}

\begin{figure*}[!t]
    \center{\includegraphics[width=1\textwidth]{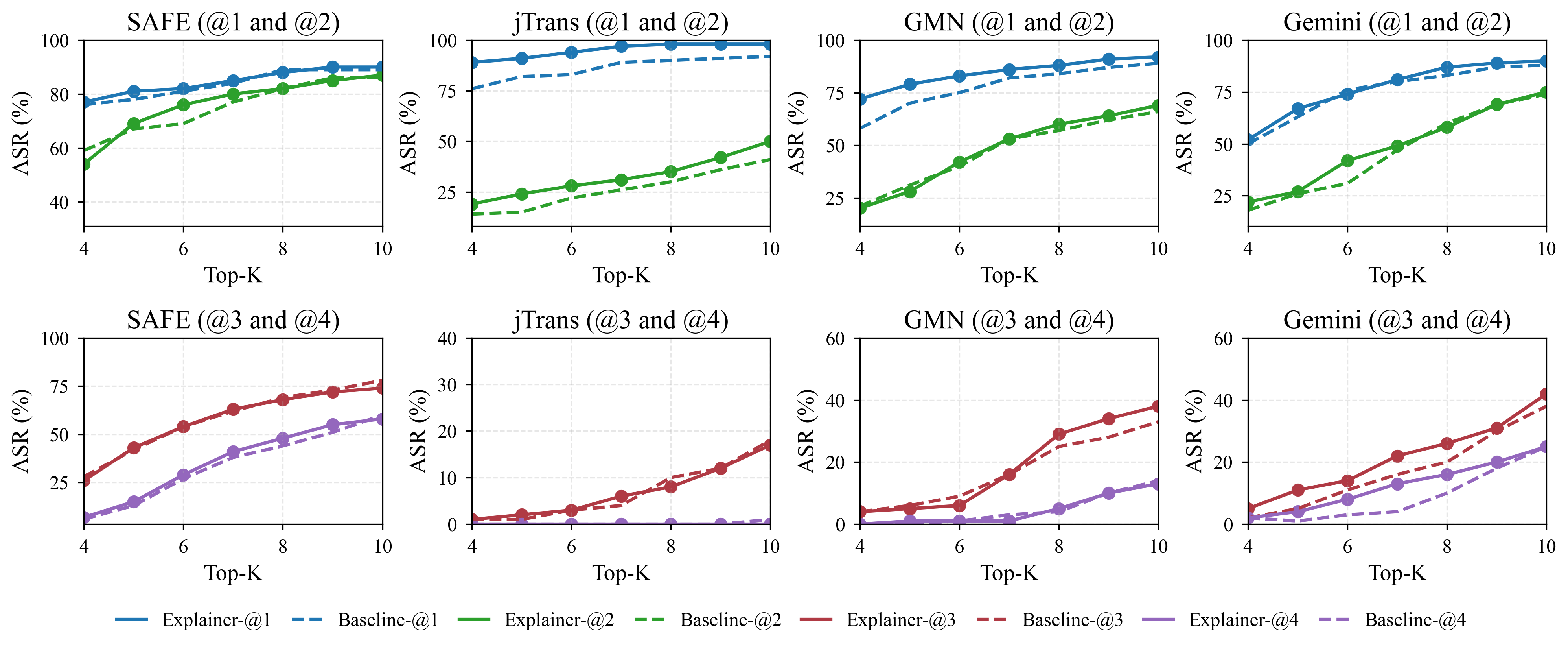}}
    \caption{ASR with varying K for our attacks, with a pool size of 128.
    }
    \label{fig:varyingk-128}
\end{figure*}

\begin{table}[h]

    \caption{Statistics about the sampled functions in our dataset. The numbers denote the average value observed across all sampled functions.}
    \centering
    {
    \begin{tabular}{l|c|c|c|c}
    \toprule
    \textbf{} & \textbf{gcc-O0} & \textbf{gcc-O3} & \textbf{clang-O0} & \textbf{clang-O3} \\
    \toprule
    \#Basic Blocks & 105.88 & 219.64 & 116.24 & 269.0 \\
    \midrule
    \#CFG Edges & 69.44 & 137.96 & 80.0 & 167.40 \\
    \midrule
    \#Instructions & 445.08 & 640.84 & 411.28 & 775.28\\
    \midrule
    Cyclic Complexity & 38.44 & 83.68 & 38.24 & 103.60 \\
    \bottomrule
    \end{tabular}
    }
    \label{tab:stats-functions}
\end{table}

\subsection{Attack Effectiveness (RQ1)}
\label{s:eval-asr}
As the real-world projects in our dataset defined over 127K functions, it would be very time-consuming, if not infeasible, to conduct the attacks using all of them.
To evaluate the effectiveness under practical complex attack scenarios, in our experiments, we randomly sampled 100 functions (25 functions under all 4 compilation configurations), with sufficient complexity across the object files in the compiled projects. 
The detailed statistics are listed in \autoref{tab:stats-functions}. 
For each function sample, we randomly selected another function in the evaluation dataset and used its four compiled variants as the target functions.
The configurations are aligned with the baseline attacks.

The attack success rates and required modifications are presented in \autoref{tab:asr-top5} with our explainer-guided attacks (denoted with a suffix +) against baseline attacks across various models.
The evaluation spans both sequence-based models and graph-based models using pool sizes of 128 and 512, and K as 5. 
The function pool was randomly selected from all functions in the object files.
The evaluation results demonstrate that our explainer-guided attacks achieved a higher success rate in almost all scenarios.
For example, in the jTrans+ model at ASR@1, we achieved a success rate of {0.91} compared to {0.78} for the baseline, indicating enhanced effectiveness when guided by explainers.
Even for more challenging goals like ASR@2 and ASR@3, the explainer-guided attacks consistently lead to higher success rates across most models.
One exceptional case is GMN, where ASR@1 and ASR@4 appeared to be higher than the baseline on a small function pool, while the ASRs were slightly lower than the baseline in other settings.
This may be attributed to GMN's iterative node aggregation algorithm, which inherently dilutes the localized impact of perturbations targeting single instructions, potentially leading to varied efficacy depending on the context.

As we used a small K in previous experiments that imposed tough constraints on the attacks, we further evaluated how varying this value may affect the ASRs.
As shown in \autoref{fig:varyingk-128}, our ASRs increased with larger K's.
In particular, the explainer-guided attacks generally achieved higher ASR@1 and ASR@2 with K=10, whereas the advantage appeared less significant in the more challenging scenarios for ASR@3 and ASR@4.

Regarding the required modifications, our explainer-guided attacks are generally comparable to or even more efficient than the baseline. %
This is particularly evident in graph-based models such as GMN+ and Gemini+, where the number of inserted instructions and nodes is often similar to or smaller than their respective baselines.
While they sometimes introduce slightly more changes (\eg{,} jTrans+@2 requires 152.0 instructions vs. 83.5 for the baseline), such cases are exceptions rather than the norm.
These results confirm that using explainers to identify salient instructions for perturbation leads to more effective and efficient attacks.
Across both model architectures, we observe that attacks conducted under a smaller model pool size of 128 tend to achieve higher attack success rates with lower modification costs.

\mybox{white}{
\textbf{Takeaway.}
Our explainer-guided attacks can achieve higher attack success rates than the baseline in almost all testing scenarios, while
requiring a comparable amount of perturbations.
}

\subsection{Attack Efficiency (RQ2)}
\label{s:eval-time}
We present the runtime overhead for selecting the important instructions and generating an adversarial sample in \autoref{tab:overhead-targeted}.
The numbers denote the average runtime overhead we observed on all function samples evaluated.
The adversarial sample (AS) generation time refers to the overall runtime overhead for constructing the adversarial sample.

As shown, the explainers effectively located important instructions with higher efficiency on all target models (\eg{,} 32.86 seconds for Gemini+ and 113.23 seconds for jTrans+).
Notably, we observed a speedup of 12.71x for instruction selection on Gemini, while the optimization for SAFE was the least impactful at 30.99\%.
On the other hand, as instruction selection only accounts for a small portion of the total execution time, the speedup on AS generation is less significant.
However, the explainer-guided attacks were still able to finish more quickly, reducing the overhead by up to 10.45\%.

\begin{table}[!h]

\caption{Time overhead for generating adversarial samples in our targeted attacks.}
\small
\centering
\resizebox{0.49\textwidth}{!}
{
\begin{tabular}{l|llllllll|llllllll}
\toprule
\multirow{2}{*}{\textbf{}}                        & \multicolumn{8}{c|}{\textbf{Sequence-based Models}}                                                                                                 & \multicolumn{8}{c}{\textbf{Graph-based Models}}                                                                                                      \\ \cmidrule{2-17} 
                                                  & \multicolumn{2}{c|}{\textbf{SAFE+}} & \multicolumn{2}{c|}{\textbf{SAFE}} & \multicolumn{2}{c|}{\textbf{jTrans+}} & \multicolumn{2}{c|}{\textbf{jTrans}} & \multicolumn{2}{c|}{\textbf{Gemini+}} & \multicolumn{2}{c|}{\textbf{Gemini}} & \multicolumn{2}{c|}{\textbf{GMN+}} & \multicolumn{2}{c}{\textbf{GMN}} \\ 
\midrule
\multicolumn{1}{l|}{\textbf{Instr Sel. (s)}} & \multicolumn{2}{l|}{101.95}               & \multicolumn{2}{l|}{133.54}              & \multicolumn{2}{l|}{113.23}               & \multicolumn{2}{l|}{286.09}              & \multicolumn{2}{l|}{32.86}                 & \multicolumn{2}{l|}{450.54}                & \multicolumn{2}{l|}{433.31}              & \multicolumn{2}{l}{598.13}             \\ 
\midrule
\multicolumn{1}{l|}{\textbf{Speedup (\%)}}         & \multicolumn{4}{c|}{30.99}                                                    & \multicolumn{4}{c|}{152.66}                                                    & \multicolumn{4}{c|}{1271.09}                                                        & \multicolumn{4}{c}{38.04}                                                  \\ 
\midrule
\textbf{AS Gen. (s)}                        & \multicolumn{2}{l|}{2735.98}               & \multicolumn{2}{l|}{2738.48}              & \multicolumn{2}{l|}{3236.16}               & \multicolumn{2}{l|}{3369.31}              & \multicolumn{2}{l|}{3519.19}                 & \multicolumn{2}{l|}{3887.07}                & \multicolumn{2}{l|}{4759.39}              & \multicolumn{2}{l}{5050.06}             \\ 
\midrule
\multicolumn{1}{l|}{\textbf{Speedup (\%)}}        & \multicolumn{4}{c|}{0.09}                                                    & \multicolumn{4}{c|}{4.11}                                                    & \multicolumn{4}{c|}{10.45}                                                        & \multicolumn{4}{c}{6.11}                                                  \\ 
\bottomrule
\end{tabular}
}
\label{tab:overhead-targeted}
\end{table}

\mybox{white}{
\textbf{Takeaway.}
Our explainer-guided attacks are more efficient by focusing on the most vulnerable code for perturbation.
Compared with the baselines, our attacks achieved a maximum speedup of 12.71x on instruction selection, while the overall overhead was reduced by up to 10.45\%.
}

\subsection{Transferability (RQ3)}
\label{s:eval-transfer}
We further investigated the transferability of attacks, \ie,
whether the adversarial samples for one BCSD model can be generalized to target other models.
We evaluated both our explainer-based attack and the baseline attack~\cite{capozzi2024lack}.

As illustrated in \autoref{tab:transferability-comparison}, when using our adversarial samples against jTrans and SAFE (target models), our attacks always exhibited superior transferability.
On the graph-based models, however, the baseline approach could be more effective, \eg, when using the sequence-based samples against graph-based models.
This could be caused by the significant architectural disparity (\ie, sequence vs. graph), which hinders the transfer of highly specialized perturbations on the salient instructions selected by explainers.
We also observed that the adversarial samples generated for GMN tend to be less successful in attacking Gemini.
One possibility is the stricter structural invariants during graph induction in Gemini
created higher curvature decision boundaries that resist out of manifold perturbations.
Nonetheless, the adversarial samples generated during our attacks in general achieved higher transferability in most tests.

\begin{table}[htbp]
    \centering
    \caption{Transferability matrix for the targeted attack case, considering $|P| = 128$ and $K = 10$. 
    In the rows, we indicate the model for which the adversarial samples were created, and in the columns, the model on which the samples were tested. Each value represents the wASR (\%).} %
    \begin{tabular}{l|l|cccc}
    \toprule
    \multicolumn{2}{c|}{} & \multicolumn{4}{c}{Target Model} \\
    \cmidrule{3-6}
    Source & Variant & jTrans & SAFE & Gemini & GMN \\
    \midrule
    \multirow{2}{*}{jTrans} & Explainer & --- & \textbf{59.50} & \textbf{39.75} & 42.50 \\
    & Baseline & --- & 59.25 & 38.75 & \textbf{47.00} \\
    \midrule
    \multirow{2}{*}{SAFE} & Explainer & \textbf{40.75} & --- & 41.00 & 44.00 \\
    & Baseline & 40.25 & --- & \textbf{42.75} & \textbf{49.00} \\
    \midrule
    \multirow{2}{*}{Gemini} & Explainer & \textbf{31.00} & \textbf{32.75} & --- & \textbf{96.00} \\
    & Baseline & 28.00 & 32.00 & --- & 95.75 \\
    \midrule
    \multirow{2}{*}{GMN} & Explainer & \textbf{30.45} & \textbf{36.00} & 65.50 & --- \\
    & Baseline & 27.75 & 33.00 & \textbf{67.75} & --- \\
    \bottomrule
    \end{tabular}
    \label{tab:transferability-comparison}
\end{table}

\mybox{white}{
\textbf{Takeaway.}
The adversarial samples generated in our explainer-guided attacks tend to be effective when used against other BCSD models in the same architecture.
The graph-based models like Gemini and GMN, which likely implement stricter decision boundaries, are more difficult to mislead using cross-model adversarial samples.
}
\subsection{Real-world Implications (RQ4)}

To assess the real-world impact, we evaluated our approach in two real-world security tasks: vulnerability detection evasion and vulnerability classification misguidance.

\subsubsection{Vulnerability Detection Evasion}
BCSD models are widely used for vulnerability detection by identifying whether an input binary contains known vulnerabilities through similarity comparison.
Vulnerability detection evasion aims to deceive these models into misclassifying a \emph{vulnerable function} as \emph{a targeted benign function}, leaving the vulnerability undetected.
To evaluate the effectiveness of our approach, we target OpenSSL~\cite{openssl}, a widely used SSL/TLS library.
Specifically, we conduct experiments on OpenSSL versions 3.0.8 and 3.3.3.
We included five recent vulnerabilities (CVE-2023-0215, CVE-2023-0216, CVE-2024-5535, CVE-2024-6119, and CVE-2024-9143), spanning various categories, including memory safety, cryptographic weaknesses, and protocol-level flaws.

To identify vulnerable functions for our attack, we first analyzed the patches for each vulnerability and selected the functions that were modified.
We constructed a pool with 128 functions using the same four compilation configurations.
In this experiment, we selected jTrans as the target model.
We measure three metrics: 
(1) \emph{initial similarity}, which is the similarity between the input samples and the target function before the attack; 
(2) \emph{final similarity}, which captures the similarity after the attack; 
and (3) \emph{average similarity}, defined as the average similarity of all samples in the pool to the target function. 
Note that the average similarity is the same before and after the attack, as the pool does not change.

As shown in \autoref{tab:detection-evasion}, our attack consistently increased the similarity scores across all the vulnerabilities.
The average initial similarity was 0.869, which is lower than the average pool similarity of 0.876. 
This indicates that the input samples were less similar to the target functions and thus more challenging to manipulate. 
After applying our attack, the average final similarity rose to 0.917, surpassing both the initial and average pool similarities.
This confirms that our method not only increases similarity but does so in a targeted and effective way. 
It makes the adversarial examples appear more functionally aligned with the target than even the average benign samples, significantly increasing the chance for evading detection.

\begin{table}[h]
    \caption{Similarity scores before and after our attack in real-world OpenSSL vulnerabilities.}
    \centering
    \resizebox{0.45\textwidth}{!}{
    \begin{tabular}{l|c|c|c}
    \toprule
    \textbf{CVE} & \textbf{Init Similarity} & \textbf{Final Similarity} & \textbf{Average Similarity} \\
    \toprule
    CVE-2024-9143 & 0.898 & 0.929 & 0.874 \\
    CVE-2024-6119 & 0.867 & 0.908 & 0.882 \\
    CVE-2024-5535 & 0.843 & 0.920 & 0.869 \\
    CVE-2023-0215 & 0.854 & 0.885 & 0.873 \\
    CVE-2023-0216 & 0.884 & 0.942 & 0.881 \\
    \midrule
    {Average} & {0.869} & {0.917} & {0.876} \\
    \bottomrule
    \end{tabular}
    }
    \label{tab:detection-evasion}
\end{table}

\subsubsection{Vulnerability Classification Misguidance}
The categorization of vulnerabilities plays a critical role in guiding the patching and mitigation process.
For instance, by grouping vulnerabilities based on the categories, developers can prioritize fixes that address multiple attack surfaces simultaneously.
Vulnerability classification misguidance aims to deceive the classification model into misjudging the severity or category of a vulnerability.

In our experiments, we selected CWEs from the CWE Most Dangerous Software Weaknesses list~\cite{cwe-top25}, with the goal of misleading the model into misclassifying code as a specific target CWE category. 
Vulnerable code samples containing these CWE types were obtained from the National Vulnerability Database~\cite{nvd} and the NIST Software Assurance Reference Dataset~\cite{sard}.
To simulate adversarial conditions, we constructed a pool of malicious functions with a size of 128, compiled under four consistent compilation settings. 
We selected CWE category pairs (original and target) with clear semantic and structural differences.
For example, we include CWE-121 (Stack-based Buffer Overflow) vs. CWE-190 (Integer Overflow), and CWE-134 (Externally-Controlled Format String) vs. CWE-193 (Off-by-one Error). 
The fundamental differences make the misclassification more challenging and meaningful.
Similarly, the experiments were conducted against jTrans, and

we measured the initial, final, and average similarity as well.

As shown in \autoref{tab:classification-evasion}, our method consistently increases the similarity scores across all targeted CWE pairs.
On average, the initial similarity was 0.859, which is lower than the average pool similarity of 0.883.
However, after applying adversarial perturbations, the final similarity rose to 0.953, exceeding both the average similarity of the CWE functions pool and the initial similarity.
This demonstrates that our attack can effectively reduce the distinction between unrelated CWE categories.
The results confirm the effectiveness of our attack strategy in vulnerability classification misguidance and highlight the risk in vulnerability management.

\begin{table}[h!]
    \caption{Similarity scores for CWE classification misguidance.}
    \centering
    \resizebox{0.47\textwidth}{!}{
    \begin{tabular}{l|c|c|c}
    \toprule
    \textbf{CWE Pair} & \textbf{Init Similarity} & \textbf{Final Similarity} & \textbf{Average Similarity} \\
    \toprule
    CWE121 - CWE190 & 0.870 & 0.977 & 0.916 \\
    CWE121 - CWE134 & 0.867 & 0.908 & 0.888 \\
    CWE190 - CWE121 & 0.869 & 0.959 & 0.924 \\
    CWE190 - CWE193 & 0.869 & 0.959 & 0.924 \\
    CWE134 - CWE190 & 0.838 & 0.971 & 0.913 \\
    CWE134 - CWE121 & 0.843 & 0.945 & 0.910 \\
     \midrule
    {Average} & {0.859} & {0.953} & {0.912} \\
    \bottomrule
    \end{tabular}
    }
    \label{tab:classification-evasion}
    \end{table}

\mybox{white}{
\textbf{Takeaway.}
Our attacks are not only effective in manipulating model outputs but also practical in compromising real-world applications of BCSD models by evading vulnerability detection and  misleading vulnerability categorization.
}

\section{Discussion}
\label{s:discussion}
We now discuss the limitations and future work.

\PP{Optimizations}
In our current implementation, the selected instructions are manipulated similarly to existing studies~\cite{capozzi2024lack, capozzi2024adversarial}.
We adopt such a design to enable fair comparison with state-of-the-art attacks.
Nonetheless, our attacks can be easily extended to incorporate more advanced code transformations.
For example, program obfuscation techniques can effectively disrupt the static disassembly process. %
Other viable methods include edge hiding ~\cite{jia2024enhancing} and code cloning~\cite{jia2024enhancing}, which aim to alternate or obscure certain control flow paths.
Junk code insertion~\cite{junk-code} involves adding instructions like \texttt{if (false) \{ jmp \}}, which causes the bytes of subsequent normal instructions to be misinterpreted as operands of the current \texttt{jmp} instruction, thereby disrupting the structure of the current instruction and leading to anomalies in static analysis.
We leave it for future work to explore more effective instruction manipulation approaches to further boost the attack performance.

\PP{Extensibility}
In this work, we focused our evaluation on four representative sequence and graph-based %
models.
Nevertheless, the explainer-guided attack principle can also be readily applied to other categories of models, including neural network based models like  BinFinder~\cite{qasem2023binary} and Zeek~\cite{shalev2018binary}, \etc{}
Additionally, we believe explainers can be applied to advance the attack against other program analysis models.
Besides, the explainer-guided attack method can also be applied to untargeted tasks.
For example, our explainer can similarly identify critical instructions and perturb them, aiming to minimize the similarity score between the perturbed (source) function and its variants.
We plan to investigate these in the future.

\PP{Feature Extraction}
Our explainer leverages the features extracted by BCSD models to guide adversarial perturbations.
This assumption is practical and often holds in real-world and research settings.
Many BCSD models rely on explicit feature extraction pipelines, such as disassembly, CFG construction, and instruction-level analysis, implemented using standard binary analysis frameworks like IDA Pro~\cite{idapro}, Binary Ninja~\cite{binaryninja}, and angr~\cite{angr}.
In practice, these intermediate features are frequently retained or exported by the toolchains for integration with other modules, \eg, for visualization or further analysis, making them accessible in most %
attack scenarios. 
Even when features are not directly available, we can often approximate them by replicating the model’s preprocessing pipeline or intercepting data at various stages, enabling our explainer to operate effectively in such cases as well. 

\PP{Defense}
Existing studies have proposed several defense against the relevant adversarial attacks.
Adversarial training approaches like FuncFooler~\cite{jia2024enhancing} enhance resilience by injecting structurally perturbed examples during training. However, such methods are tightly coupled to the observed perturbation patterns, and may not generalize to novel attacks like ours. 
Other techniques, such as GWAD~\cite{park2025mind}, attempt to detect adversarial behaviors by analyzing the query dynamics.
While effective in certain scenarios, these methods are often limited by their reliance on predefined detection heuristics, making them less robust against sophisticated attacks that bypass such patterns.
To effectively defend against our explainer-guided attacks, one possible way is to reduce the precision of explanation, \eg, using gradient obfuscation, attention randomization, and saliency regularization~\cite{yue2023gradient}. 
Input-based defense like (De)Randomized Smoothing~\cite{salem2024robust} may also dilute the impact of localized perturbations, increasing the robustness against crafted perturbations. 
We believe more efforts need to be invested to explore the possible defense.

\section{Related Work}
\label{s:relwk}

\PP{Adversarial Attacks against Source Code Analysis}
Various advanced models for source code analysis have been proposed, effectively improving the performance in tasks like clone detection, method name prediction, \etc~Consequently, extensive efforts have been invested to measure the robustness of such models via adversarial attacks.
Yefet \etal~\cite{yefet2020adversarial} designed a white-box adversarial attack against code models for Java and C\# analysis. They assume adversaries have the knowledge about the gradients of targeted models.
Zhang \etal~\cite{zhang2023challenging} implemented fifteen semantic-preserving code transformation for constructing adversarial samples against machine learning-based clone detection models.
Quiring \etal~\cite{quiring2019misleading} focused on semantic equivalent coding style transformation, dramatically downgrades the accuracy of authorship attribution models.
Other affected application scenarios include plagirism detection, where genetic code transformation techniques were implemented to undermine the detection models~\cite{devore2020mossad}.
The above research appears orthogonal to this work, which focuses particularly on binary code similarity detection models.

\PP{Attacks against Binary Analysis Models}
Binary code analysis models have also been a popular target for various attacks.
Capozzi \etal ~\cite{capozzi2024lack} adopted a brute-force strategy for selecting instructions to perturb.
Specifically, they exhaustively traversed and removed each assembly instruction, and measured the importance of instructions according to the similarity score changes.
They also selected the code transformations for each candidate instruction based on the similarity score changes, implementing both black-box and white-box attacks against three state-of-the-art BCSD models~\cite{capozzi2024adversarial}.
Song \etal~\cite{song2022mab} focused particularly on malware classifiers and aimed to generate adversarial samples to evade malware detection and analysis.
Jia \etal~\cite{jia2022funcfooler} selected candidate instructions to mutate based on the heuristic rule that instructions in basic blocks that locate on all paths from the entry to exit must be important.
Kreuk \etal~\cite{kreuk2018adversarial} created non-executable code sections for appending the adversarial bytes, thereby preserving the original functionalities while evading the malware detection.
Similar strategies were also applied in~\cite{kolosnjaji2018adversarial}.
In contrast, Lucas \etal~\cite{lucas2021malware} proposed to mutate functional instructions, and iteratively optimize the attack effects by measuring the misclassification probabilities.
Besides adversarial attacks, backdoor vulnerabilities in binary code analysis models have also been investigated~\cite{zhang2023pelican}.

In this work, we explored another viable direction to optimize the attack effectiveness, \ie, using explainers to pinpoint important instructions for manipulation, which can be integrated with existing attacks to further boost the performance.

\PP{Explainer-guided Program Analysis and Attacks}
Explainers have been applied in other program analysis and attacks.
He \etal~\cite{he2023finer,he2022msdroid} generated  explanations for Android malware detection models to enhance the usability.
Arp \etal~\cite{arp2014drebin} augmented Android malware detectors by calculating the importance score of each feature and constructing explanations for the detection results accordingly.
In addition to generating explainable results, explainers are also used to facilitate other attacks, \eg, backdoor attacks~\cite{severi2021explanation}.
In this work, we demonstrated that explainers can also effectively optimize the adversarial attacks against binary similarity detection models.
\section{Conclusion}
\label{s:conclusion}

In this work, we introduced an optimization for targeted adversarial attacks against BCSD models.
By leveraging off-the-shelf explainers to pinpoint the salient instructions for perturbation, we are able to generate effective adversarial function samples in a computationally efficient way.
The evaluation on binary functions from real-world projects proves that explainers provide actionable and granular guidance for adversarial manipulation, significantly improving attack efficacy. 
The discoveries highlight the lack of robustness in existing BCSD models, demonstrating the possibility to hinder vulnerability detection and classification in practice.
We further emphasize the necessity of further research to enhance the robustness of BCSD models against adversarial attacks, particularly through the development of defense mechanisms that address the exploitability of explanation-driven weaknesses.

\balance
\bibliographystyle{IEEEtran}

\footnotesize
\bibliography{p,lab,conf}
\normalsize

\end{document}